\documentclass{jpp}

\newcommand{\beq}{\begin{eqnarray}}
\newcommand{\eeq}{\end{eqnarray}}

\newcommand{\ddd}[2]{\frac{\partial #1}{\partial #2}}

\newcommand{\vp}{{v_\perp}}

\newcommand{\vv}{{\bf v}}
\newcommand{\vl}{{v_\parallel}}

\def\hblabel#1{\label{#1}}
\newcommand{\x}{{\bf x}}
\newcommand{\vz}{v_{\parallel}}
\newcommand{\fd}[2]{\frac{\displaystyle #1}{\displaystyle #2}}
\newcommand{\cE}{{\cal E}}
\newcommand{\cEz}{{\cE_\parallel}}
\newcommand{\cEp}{{\cE_\perp}}

\def\eq#1{(\ref{eq:#1})}
\def\ddt#1#2{\partial #1 /  \partial #2}

\newcommand{\aveC}[1]{\left< #1 \right>_{\xi}}
\newcommand{\aveP}[1]{\left< #1 \right>_{\parallel}}
\newcommand\nuP{\nu_{\parallel}}
\newcommand\nuC{\nu_{\xi}}

\usepackage[utf8]{inputenc}
\usepackage[T1]{fontenc}

\usepackage{graphicx}
\usepackage{amsmath}
\usepackage{amssymb}
\usepackage{times}
\usepackage{bm}
\usepackage[square]{natbib}

\usepackage{multirow}
\usepackage{hhline}
\usepackage[table,dvipsnames]{xcolor}
\usepackage{rotating}
\usepackage{array}
\usepackage{textcomp}
\usepackage{float}
\usepackage{upgreek}

\definecolor{blue}{rgb}{0,0,1}
\usepackage[colorlinks=true,pdfborder={0 0 0},allcolors=blue]{hyperref}

\graphicspath{{figs/}}

\title{ Parallel Velocity Mixing Yielding Enhanced Electron Heating During  Magnetic Pumping} 

\author{J.Egedal\aff{1} \corresp{\email{egedal@wisc.edu}},
J. Schroeder\aff{1}, E. Lichko\aff{2}  }
\affiliation{\aff{1}Department of Physics, University of Wisconsin-Madison, Madison, Wisconsin,  USA
\aff{2}Lunar and Planetary Laboratory, University of Arizona, Tucson, AZ, USA}

\begin{document}

\maketitle

\begin{abstract}
Magnetic wave perturbations are observed in the solar wind and in the vicinity of Earth's bow shock. For such environments, recent work on magnetic pumping with electrons trapped in the magnetic perturbations have demonstrated the possibility of efficient energization of superthermal electrons. Here we also analyze the energization of such energetic electrons for which the transit time through the system is short compared to time scales associated with the magnetic field evolution. In particular, considering an idealized magnetic configuration we show how trapping/detrapping of energetic magnetized electrons can cause effective parallel velocity ($\vz$-) diffusion. This parallel diffusion, combined with naturally occurring mechanisms known to cause pitch angle scattering, such as Whistler waves, produces enhanced heating rates for magnetic pumping.  We find that at low pitch angle scattering rates the combined mechanism enhances the heating beyond the predictions of the recent theory for magnetic pumping with trapped electrons.
\end{abstract}

\maketitle

\section{Introduction}

The transport of matter and radiation in the solar wind and terrestrial magnetosphere is a complicated problem involving competing processes of charged particles interacting with electric and magnetic fields. Given the rapid expansion of the solar wind within the Parker spiral, it would be expected that superthermal particles originating in the corona would cool rapidly as a function of distance to the Sun. However, observations show that this is not the case and superthermal particles have been observed out to the termination shock \cite{decker:2008}, suggesting the presence of an additional heating/acceleration mechanism. These superthermal tails have been observed to follow a power-law distribution in velocity space \cite{fisk:2006}.

Much of the work on a possible explanation for this additional heating centers on wave-particle interactions as the primary heating mechanism, where energy is provided by the turbulence associated with propagating waves \cite{kennel:1966,ergun:1998,vinas:2000,rhee:2006,califano:2008,vocks:2005}. In these models, particles are energized at the resonant velocities, where $v k \cos\Theta\simeq \omega$, and with $\cos\Theta = \vv\cdot{\bf k}/(vk)$. Particle energization is then limited to $v\leq \omega/(k\cos\Theta)$. Superthermal electrons then require energization by waves with large phase velocities $v_p=\omega/k$, such as Whistler waves \cite{wilson:2012}. However, in many systems, the energy available in Whistler waves has been found to be insufficient to explain the observed level of electron energization and in a recent analysis using MMS data it was found that while whistlers are effective for pitch angle scattering, the whistler bursts did not correlate well with electron energization \cite{oka:2017}. 

Another challenge in using wave-particle interactions to explain the heating in the solar wind is the near-ubiquitous observations of power-law distributions of superthermal particles. Such power-law distributions are known to form in Fermi-like heating processes where the energy gains of individual particles are proportional to their initial energies, but it is difficult to reproduce with a set of resonant wave-particle interactions.

Previous work on magnetic pumping (such as  transit-time damping \cite{berger:1958,barnes:1966,stix:1992,lichko:2017}), have largely concluded that pumping is not efficient for energizing superthermal electrons with $v>\omega/(k \cos\Theta)$.
 Meanwhile, the pumping models of Ref.~\cite{egedal:2018,lichko:2020a} include the effects of  trapping and differs significantly from earlier results, as the trapping permits especially electrons with   $v\gg\omega/(k \cos\Theta)$ to become energized. Another interesting property that is in contrast to the turbulent cascade where the energy is transferred from large scales to small scales before being absorbed \cite{sahraoui:2009,howes:2008}, in magnetic pumping the energy is provided directly by the energy rich largest scale magnetic fluctuations.


The results of the present paper can be considered an extension of the magnetic pumping model by \cite{lichko:2020a}. Here we uncover an additional heating mechanism, which is related to the particular effect of electrons becoming trapped/untrapped in magnetic mirror-structures that form in the presence of compressional wave dynamics. The process leads to parallel energy mixing (or $\vz$-mixing) yielding a net energy gain also for electrons for which the magnetic moment, $\mu$, is conserved. In turn, by adding weak pitch angle scattering to the system a heating model is obtained similar to that of \cite{lichko:2020a}, with the main difference being enhanced heating rates of superthermal electrons in the limit of weak pitch angle scattering. 

The paper is organized as follows: In Section II we evaluate the parallel electron energization and mixing in  a system of magnetic trapping/detrapping where the magnetic moment, $\mu$ is consider an adiabatic invariant. In the drift-kinetic limit and the limit of fast orbit bounce motion, in Section III we show how this $\vz$-mixing leads to parallel diffusion within the trapped part of the electron distributions. In Section IV we add to the model a phenomenological pitch angle scattering and evaluate the net changes to a distribution which after complete $\vz$-diffusion becomes reisotropized by the pitch angle scattering. We then in Section V  consider a scenario of simultaneous $\vz$-mixing and pitch angle diffusion, for which we derive an evolution equation for the slowly varying background electron distribution. In Section VI the new results are discussed and the analysis concluded with a comparison to those of \cite{lichko:2020a}, emphasizing enhanced heating rates for low values of pitch angle scattering.

\section{Parallel Energy Mixing }

Waves including magnetic perturbations can trap electrons, and considering an idealized standing wave configuration \cite{lichko:2020a} demonstrated the importance of trapping to render magnetic pumping
an efficient heating mechanism for superthermal electrons. In general, however, magnetic perturbations will have a range of wave-lengths, amplitudes and phases such that at different locations along a magnetic flux-tube regions of trapped electrons can develop and interact in a range of ways not accounted for in the previous analysis. We here explore how the process of trapping/detapping itself leads to parallel energy mixing ($\vz$-diffusion), with the result of heating even for the case where the electron magnetic moments are conserved. The electron dynamics is here well accounted for by the drift-kinetic framework pioneered by  \cite{kulsrud:1983}. In this  framework the change of the electron energy is described by $\ddt{\cE}{t}= \mu\ddt{B}{t} - e({\bf v}_\|+ {\bf v}_D)\cdot{\bf E}$, where $
{\bf v}_\|$ and ${\bf v}_D$ are the field-aligned parallel streaming and guiding center drift, respectively.

As a simplifying assumption and similar to \cite{montag:2017,egedal:2018,lichko:2020a}, in present analysis we will only consider the electron dynamics in the fast transit time limit. In this limit it is assumed that the timescale associated with the electron bounce motion, $\tau_b$, is much shorter than the timescales characterizing the evolution of the magnetic perturbations, such that both the magnetic moment $\mu$, and the parallel action integral $J=\oint \vz dl$ become adiabatic invariants. In addition, we assume a 1D spatial geometry where the electrons are confined in a single flux-tube. Orbits of electrons are then fully characterized by $\mu$ and $J$, and given the fast transit-time limit we can apply the multiple timescale method \cite{davidson:1972}, where $f(\x,\vv)$ is approximately constant along the ``instantaneous'' bounce orbits. Consistent with Jean's theorem \cite{jeans:1915}, it then follows that distributions can be expressed on the form $f=g(\mu,J)$, where $g$ is an arbitrary function.  

For the present analysis, however, we find it more useful to write 
$f(\mu,\cE,t)=f_0(\mu,\cE_0)$, where $\cE_0$ is the initial particle energy at  a time $t_0$ of an initial known distribution, $f_0$. The problem of solving for the distribution $f(\mu,\cE,t)$ is then reduced to obtaining a mapping between $\cE(t)$ and $\cE_0$ consistent with conservation of $\mu$ and $J$. Because we will only consider prescribed magnetic perturbations there are no feedback of $f(\mu,\cE,t)$ onto the wave dynamics. Determining the mapping $\cE(t)\rightarrow \cE_0$ then becomes a ``single particle'' problem which can be solved by basically considering one point in phase-space, $(\mu,\cE,t)$, at a time. For general magnetic perturbations, determining the mapping $\cE(t)\rightarrow \cE_0$ is then a problem well suited for numerical  orbit integration methods. Here, however, we will consider particularly simple magnetic geometries that allow explicit expressions for $\cE_0(\mu,\cE,t)$ to be determined, which (as we will see below) then in turn provides an explicit solution for the distribution function, $f(\mu,\cE,t)=f_0(\mu,\cE_0(\mu,\cE,t))$.

\begin{figure}
	\centering
	\includegraphics[width=13 cm]{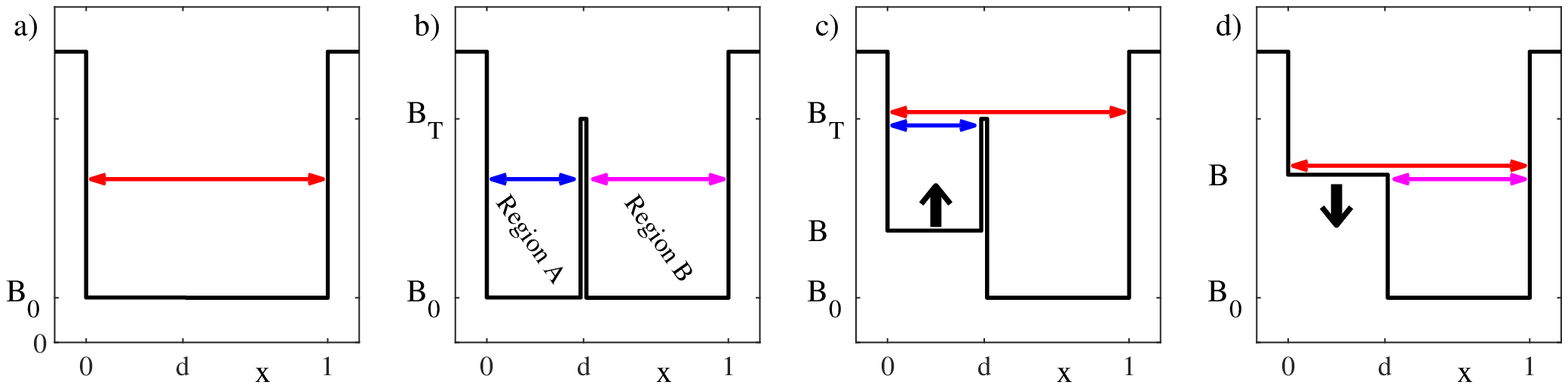}%
	\caption{ Sequence of magnetic perturbations considered for parallel velocity mixing, with the colored arrows indicating distinct trapped orbit types. The  deep magnetic square well of a) is in b) modified by a spatially narrow magnetic barrier at $x=d$ with $B=B_T$, separating Regions A and B. In c) the floor of Region A is enhanced until $B=B_T$ is reach. In d) the floor of Region A (and the barrier) is reduce bringing the configuration back the the initial state in a).
	}
	\label{fig:DoubleFig3}
\end{figure}

To illustrate how $\vz$-diffusion can occur for adiabatic electrons with  fixed magnetic moments, in Fig.~\ref{fig:DoubleFig3} we  consider a magnetic flux-tube with a square-shaped magnetic perturbation characterized by a reduced magnetic field $B_0$. Again, throughout the analysis we will assume that the electron transit time is fast compared to the timescale at which the magnetic field is changing. Thus, the evolution of the electron population will be adiabatic and reversible.  As illustrated in Fig.~\ref{fig:DoubleFig3}(a), the magnetic well can trap electrons. The locations inside the magnetic well are parameterized by $x\in [0; 1]$, and at $x=d$ a narrow region is introduced where the magnetic field is increased to an enhanced value $B_T$. The width of this enhancement is assumed to be so narrow that we can neglect any heating $\mu\ddt{B}{t}$ that result as it builds in time. Thus, the role of $B_T$ is solely to split electron orbits with total energy $\cE< \mu B_T$ into locally trapped orbits in regions-A and -B, indicated in Fig.~\ref{fig:DoubleFig3}(b). 

From this point the magnetic geometry can be modified in a variety of ways. For example, in appendix A, we analyse the result of changing dynamically the location $d$ of the barrier field $B_T$, and obtain very similar results to those to be derived now for a very different magnetic evolution. In this main section we consider a mixing cycle where we slowly raise the magnetic field in region-A until it reaches the value of the barrier, $B_T$. During this process  all the electrons trapped in region-A will be energized at a rate $\mu \ddt{B}{t}$ and become un-trapped as their total energies reach $\cE=\mu B_T$. Here the electron orbits will undergo a transition from the blue orbit type  to the red orbit type  in Fig.~\ref{fig:DoubleFig3}(c). In region-B the magnetic field is constant and no heating occurs, and compared to the trapped orbits in region-A, the red orbit types are subject to reduced heating rates related the reduced fraction of time a given electron is present in region-A. Then, as illustrated in  Fig.~\ref{fig:DoubleFig3}(d) we again lower the magnetic field in region-A and also eliminate the previous magnetic barrier at $x=d$. During this process, all the electrons  initially confined to  region-B will observe orbit changes  corresponding to the transition from the magenta to the red orbit types in Fig.~\ref{fig:DoubleFig3}(d).  All electrons traversing regions-A and -B will be cooled  proportionally the relative faction of the time they spend in region-A.

As indicated above, the heating of the various orbit types can in principle  be computed by integrating  the  orbit averaged value of   $\mu\ddt{B}{t}$. Meanwhile, the task of evaluating the energy changes is significantly simplified by considering  the parallel action integral
\begin{equation}
\label{Jact}
J=\sqrt{\frac{m}{2}} \oint v_\| dl \quad.
\end{equation}
Here the integral is taken over the trapped orbits' bounce motion, and the unimportant factor $\sqrt{m/2}$ is included to  ease the notation below. In addition, the use of $J$ makes the analysis more general as this framework also applies to less idealized configurations where the $ - e({\bf v}_\|+ {\bf v}_D)\cdot{\bf E}$-term (mentioned above) becomes important to the energization process \cite{montag:2017}. 

For orbits bouncing through both region-A and region-B we may differentiate between the contributions from the two regions
\[
J=J_A + J_B \quad,
\]
where
\[
J_A=\sqrt{\frac{m}{2}} \int_0^{d} v_\| dl \,,\quad
J_B=\sqrt{\frac{m}{2}} \int_{d}^1 v_\| dl\quad.
\]
Introducing the present energy $\cE$ and initial energy $\cE_0$ of an electron in the configuration, using $v_\|=\sqrt{2/m}\,\sqrt{\cE-\mu B}$, the present and initial values of these action contributions are trivially evaluated as 
\[
J_A=d\sqrt{\cE-\mu B}\,,\quad J_{A0}=d\sqrt{\cE_0-\mu B_0}\quad,
\]
and
\[
J_B=(1-d)\sqrt{\cE-\mu B}\,,\quad J_{B0}=(1-d)\sqrt{\cE_0-\mu B_0}\quad.
\]
In general, the action integrals will be conserved while the magnetic configuration is evolving slowly in time. 
An exception to this occurs during the orbit transition in  Fig.~\ref{fig:DoubleFig3}(c), where a new contribution, $J_{BT}$, from region-B is acquired. Because the energy of a newly transitioned electron is $\cE=\mu B_T$, in region-B the parallel energy will be $\mu(B_T-B_0)$ and we find
\[
J_{BT}=(1-d)\sqrt{\mu (B_T-B_0)}\quad.
\]
It then follows that electrons initially trapped in region-A, will after their transition be characterized by
\[
J_A+J_B=J_{A0} + J_{BT}\quad,
\]
from which we can relate the initial energy to the present energy 
\begin{equation}
\label{eq:dE1}
\cE_0 = \frac{1}{d^2}\left[  (1-d) \left(\sqrt{\cE- \mu B_0}-\sqrt{\mu(B_T-B_0)}  \right)
+
 d\sqrt{\cE-\mu B} \right]^2 + \mu B_0 \quad.
\end{equation}
While the electrons are confined in region-A they experience the full heating provided by $\mu \ddt{B}{t}$. This heating is stronger than the average cooling they  observe when they transit both region-A and region-B and $B$ is decreasing in region-A.
Consistently, from Eq.~\eq{dE1} it is readily shown that $\cE_0<\cE$ and it follows that  all electrons originally in region-A will gain energy during the mixing sequence.

We may consider the orbit transition in Fig.~\ref{fig:DoubleFig3}(d) where orbits confined to region-B transition into orbits passing through both regions. This transition is different from that described above (where electrons cleared a barrier $\mu B_T$ and fell into a region of lower magnetic field $B_0<B_T$, yielding a parallel energy boost). In the present transition there is no localized barrier, and the newly transitioned electrons will have $v_\|\simeq0$ during their initial traversal of region-A, and there are no  new contribution to the action integral. Thus, for the electrons transitioning out of region-B we have 
\[
J_A + J_B = J_{B0}\quad,
\]
such that 
\begin{equation}
\label{eq:dE2}
\cE_0 = \frac{ \left[\,d\sqrt{\cE-\mu B} + (1-d) \sqrt{\cE- \mu B_0}\,\right]^2}{(1-d)^2}
+ \mu B_0\,\,.
\end{equation}
From Eq.~\eq{dE2} it can be shown that $\cE_0>\cE$ and all electrons which have undergone a transition out of region-B will observe a net cooling. This is consistent with our expectation because these electrons only reach region-A during the period where $\ddt{B}{t}$ is negative. 
 
 With Eqs.~\ref{eq:dE1} and \ref{eq:dE2} we have obtained expressions for the initial energy as a function of the present energy for the orbits which have undergone orbit transitions. Similar expressions are also readily obtained for the orbits which have not undergone transitions simply by imposing $J_A=J_{A0}$, $J_B=J_{B0}$, and 
 $J_A+ J_B=J_{A0} +J_{B0}$, for the ``blue'', ``magenta'', and ``red'' orbit types, respectively. Thus, the application of the action integral permits a very effective evaluation of $\cE_0$ as a function of $\cE$ for all orbit classes continuous during the mixing sequence.

\begin{figure}
	\centering
	\includegraphics[width=13 cm]{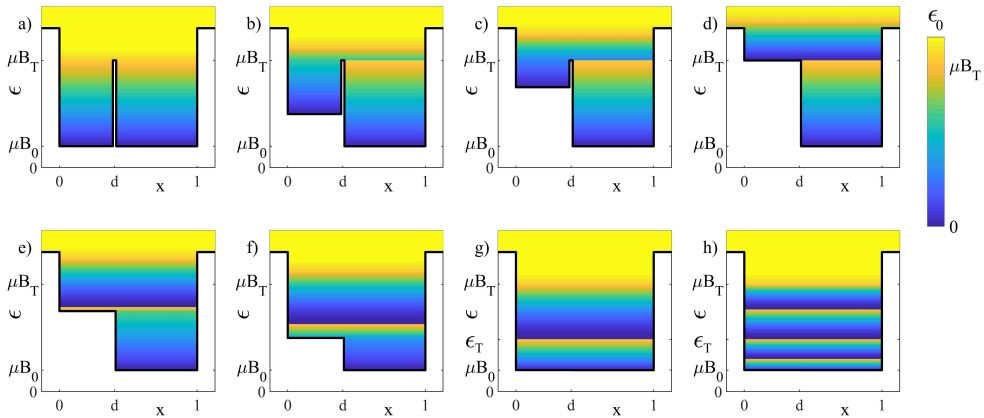}%
	\caption{a-g) Color contour plots of the initial energy $\cE_0$ as functions of position $x$ and present energy $\cE$ as observed during the evolution of the magnetic well outlined in Fig.~\ref{fig:DoubleFig3} with $d=0.4$ and $B_T/B_0=8$. Panel h) illustrates the results of two complete mixing cycles.
	}
	\label{fig:DoubleFig4}
\end{figure}

In Fig.~\ref{fig:DoubleFig4} we illustrate the evolution of $\cE_0(\cE)$ at selected times during the mixing sequence.
In this figure the $y$-axes represent the present total kinetic energy $\cE$ of  electrons with magnetic moments $\mu$, while the color contours describe their spectrum of initial energies $\cE_0$.  Fig.~\ref{fig:DoubleFig4}(a) displays the initial range of relevant energies, where we simply have $\cE=\cE_0$. The black lines are the energy barriers due to the imposed magnetic field structure. Those electrons with present energy $\cE >\mu B_T$ will be able to overcome the barrier separating region $A$ and region $B$. The changes in color   from one panel to the next then describes the evolution of the relationship between $\cE$ and $\cE_0$, as expressed in the derived equations.  Because $\mu$ is conserved during the whole mixing process, and both $\cE$ and $\cE_0$ are proportional to $\mu$, the results of the figure become applicable to any value of $\mu$.  At the time of panel (d) the magnetic field in region-A has reached that of the barrier $B=B_T$, and all the electrons originally in region-A now have energies larger than those of region-B. In panels (e,f) the magnetic field of region-A is reduced again and the magnetic barrier at $x=d$ is eliminated. 

In Fig.~\ref{fig:DoubleFig4}(g) the magnetic field has returned to its initial state. 
Consider an electron originally  marginally trapped in region-B with $\cE_0=\mu B_T$, using  Eq.~\ref{eq:dE2} with $B=B_0$ we can solve for $\cE$ to obtain the transition energy between the two electron populations after the mixing cycle is complete:
\[
\cE_T= (1-d)^2 \mu(B_T-B_0) + \mu B_0\quad.
\]
Further, using Eqs.~\ref{eq:dE1} and \ref{eq:dE2} with $B=B_0$ we obtain the mapping between $\cE_0$ and $\cE$ after one complete mixing cycle
\begin{equation}
\label{eq:dE3}
\cE_0 =  \begin{cases} \,\, \fd{\cE-\mu B_0}{(1-d)^2} +\mu B_0  & \mbox{for}\quad \mu B_0< \cE< \cE_T    \\[3ex]
\,\, \fd{1}{d^2}\left[ \sqrt{\cE- \mu B_0} - (1-d) \sqrt{\mu(B_T-B_0)}  \right]^2 + \mu B_0
 & \mbox{for}\quad  \cE_T<\cE< \mu B_T 
\\[3ex]
\,\,\cE & \mbox{for}\quad  \mu B_T< \cE
\end{cases} \,\,.
\end{equation}

As is evident from Eq.~\eq{dE3} and Fig.~\ref{fig:DoubleFig4} the energy gain $\Delta\cE=\cE-\cE_0$ depends on the initial $\cE_0$ as well as the initial location of the electrons. The electrons which gain the most energy are initially located in region-A with vanishing parallel energy, such that $\cE_0=\mu B_0$. After one mixing cycle these electrons will then have a total energy of $\cE=\cE_T$, marked in Fig.~\ref{fig:DoubleFig4}(g). On the other hand, the electrons that will be cooled the most are originally barely trapped in region~B ({\sl i.e.} $\cE_0\sim \mu B_T$). After the mixing cycle, these will also have a present energy $\cE=\cE_T$. Meanwhile, region-A electrons with $\cE_0=\mu B_T$ as well as region-B electrons with $\cE_0=\mu B_0$ observe no chance in their energies.

To characterize the effect of multiple mixing cycles we can evaluate Eq.~\ref{eq:dE3} recursively. We introduce $\cE_N$ as the electron energy after $N$  cycles and Eq.~\ref{eq:dE3} then implies that 
\begin{equation}
\label{eq:dEN}
\cE_{N-1} =  \begin{cases} \,\, \fd{\cE_N-\mu B_0}{(1-d)^2} +\mu B_0  & \mbox{for}\quad \mu B_0< \cE_N< \cE_T    \\[3ex]
\,\,  \fd{1}{d^2}\left[ \sqrt{\cE_N- \mu B_0} - (1-d) \sqrt{\mu(B_T-B_0)}  \right]^2 + \mu B_0
& \mbox{for}\quad  \cE_T<\cE_N< \mu B_T 
\\[3ex]
\,\,\cE_N & \mbox{for}\quad  \mu B_T< \cE_N
\end{cases} \,\,.
\end{equation}
With Eq.~\ref{eq:dEN} we have now established a direct mapping between the energy $\cE_N$ after $N$ mixing cycles and the initial energy $\cE_0$. An example of $\cE_0(\cE_N)$ with $N=2$ is given in  Fig.~\ref{fig:DoubleFig4}(h).

\section{Parallel diffusion of $f_e$}

\begin{figure}
	\centering
    \includegraphics[width=8.5 cm]{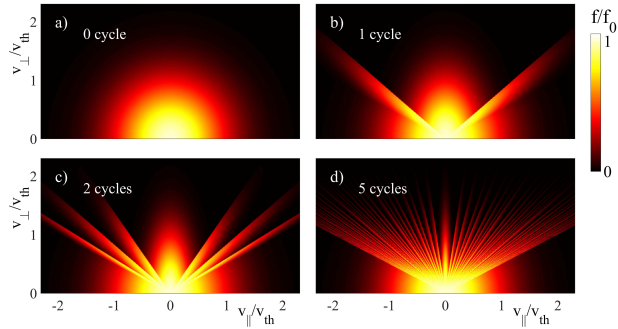}%
	\caption{For the initial distribution in a), the distributions resulting from 1, 2 and 5 mixing cycles are shown in panels b-d), respectively. The distributions are calculated using Eqs.~\ref{eq:dEN} and \ref{eq:fe} with $d=0.4$ and $B_T/B_0=8$.
	}
	\label{fig:Double}
\end{figure}

As discussed in Section II, the electrons are governed by the drift-kinetic equation, which, for the considered limit of slow magnetic field evolution and well magnetized electrons  (such that $d\mu/dt=0$), simply takes the form $df_e(\cE,\mu)/dt=0$ \cite{montag:2017}. Assuming an initial electron distribution $f_{e0}(\cE,\mu)$, with Eq.~\ref{eq:dEN} we then obtain the distribution that results after $N$ cycles as
\begin{equation}
	\label{eq:fe}
	f_e(\cE_N,\mu)= f_{e0}(\cE_0(\cE_N),\mu) \quad, 		
	\end{equation}
where $\cE_0(\cE_N)$ can be obtained from the recursion relation given in Eq.~\ref{eq:dEN}. As an example, starting with an initial Maxwellian $f_{e0}(\cE_0,\mu)$ shown in Fig.~\ref{fig:Double}(a), the results of 1, 2 and 5 mixing cycles are illustrated in Figs.~\ref{fig:Double}(b-d), respectively. 

The changes in $f_e$ induced by the parallel mixing are fully reversible. However, we note how the number of stripes in the $f_e$ grows like $2^N$ such that at sufficiently large $N$ the smallest amount of scattering will be sufficient to smooth out the exponentially narrowing  stripes. This will render $f_e$ independent of $\cE_\|$, such that for the trapped ranges affected by the pumping we have $f_e=f_e(\mu)$. At this point the mixing process has run its course and no further changes will occur in $f_e$ by parallel mixing alone. 

While the effect of parallel mixing above was calculated for a highly idealized magnetic geometry it is clear that the cause of the mixing of $f_e$ is the orbit transitions of the type introduced with Fig.~\ref{fig:DoubleFig3}(c). Therefore, any wave activity that leads to similar orbit transitions will cause  equivalent mixing in naturally occurring systems.

\section{Changes in $f_e$ due to combined $\cEz$ and pitch angle mixing }

\begin{figure}
	\centering
	\includegraphics[width=13.5 cm]{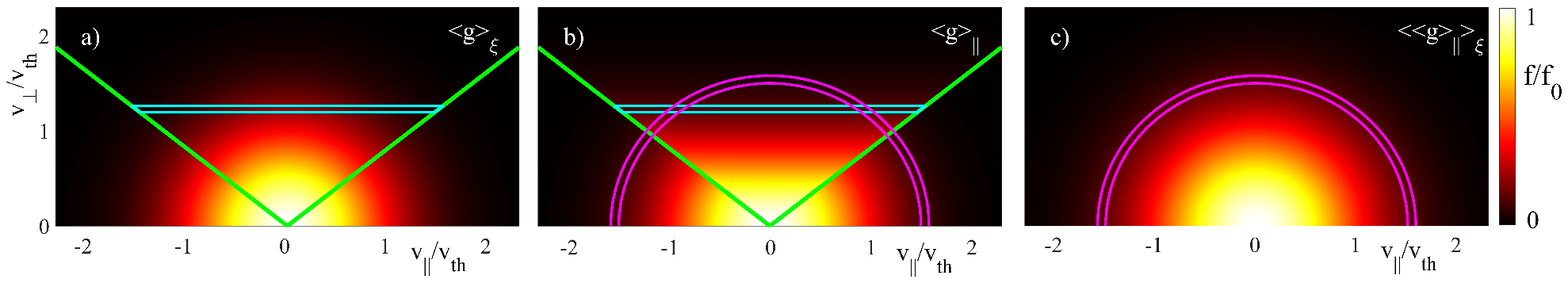}%
	\caption{Illustration of how the 2D distributions $\aveP{g}$ and $\aveC{\aveP{g}}$ are determined from $\aveC{g}$.
		In a) and b) the green lines are the trapped passing boundaries characterized by $\cEz=h\cEp$, where $h=(B_T/B_0-1)$.  The $\cEz$-mixed distribution $\aveP{g}$ is determined from  $\aveC{g}$ by requiring particle conservation for the velocity phase-space elements of the type encircled in cyan. In turn, $\aveC{\aveP{g}}$ in c) is determined from  $\aveP{g}$ by requiring particle conservation for the velocity phase-space elements of the type encircled in red in panels b) and c).
	}
	\label{fig:DoF2}
\end{figure}

In the following section we will derive a model for the heating that occurs when pitch angle scattering is included during the continuous mixing described above. We will formulate this model in terms of a slowly evolving 1D velocity distribution $g(v,t)$. 
Any distribution as a function of speed can be written as an isotropic distribution in $(v_{\perp}, v_\|)$, and we denote an initial 2D distribution as $f(v_{\perp}, v_\|)\equiv \aveC{g}$, and example of which is shown in Fig.~\ref{fig:DoF2}(a) for the case where $g(v)$ is a simple Maxwellian. Furthermore, in our 
 manipulations we will also use $\aveC{...}$ as an operator, which for any 2D distribution yields a distribution fully scattered in the cosine-pitch-angle  variable $\xi= \vl/\sqrt{\vl^2+\vp^2}$.

The distribution in Fig.~\ref{fig:DoF2}(b) represents the result of the $\vl$-diffusion described above for electrons trapped by $B_T$. We denote this distribution as $\aveP{g}$ corresponding to a distribution completely  mixed in the $\vl$-direction for the electrons within the trapped region. Here the trapped region is outlined by the green lines characterized by $\vl^2< h \vp^2$, where $h = (B_T/B_0-1)$ and $B_T$ is the value of the barrier introduced in Fig.~\ref{fig:DoubleFig3}(b).
Mathematically, $\aveP{g}$  is obtained from $\aveC{g}$ by particle conservation. In particular, we require that for any $\vp$ the rectangular type areas in the trapped regions of differential width $d\vp$, as outlined by the areas encircled in cyan in  Figs.~\ref{fig:DoF2}(a) and \ref{fig:DoF2}(b), $\aveP{g}$ and $\aveC{g}$ contain identical number of particles. 

We next consider the scenario where  $\aveP{g}$ is completely isotropized in pitch angle yielding the distribution here denoted $\aveC{\aveP{g}}$. Mathematically, as outlined in Figs.~\ref{fig:DoF2}(b) and \ref{fig:DoF2}(c) this distribution is also determined by imposing particle conservation, this time requiring  that for any $v$ the differential speed elements $dv$, as outlined by the areas encircled in magenta in Figs.~\ref{fig:DoF2}(b) and \ref{fig:DoF2}(c), contain identical number of particles in $\aveP{g}$ and $\aveC{\aveP{g}}$. In Appendix B we  show that
\begin{equation}
\hblabel{eq:pump}
\aveC{\aveP{g}}\simeq \aveC{g+ \delta g}\quad,
\end{equation}
where 
\begin{equation}
\hblabel{eq:dg}
\delta g =
\frac{h}{45 }
\left(\frac{h}{1+h} \right)^{3/2} \frac{1}{v^2}\ddd{}{v} v^4  \ddd{}{v} g \quad,
\end{equation}
and, repeated for convenience, $h=B_T/B_0-1$.

\section{Evolution of the Background Distribution }
\begin{figure}
	\centering
    \includegraphics[width=7.5 cm]{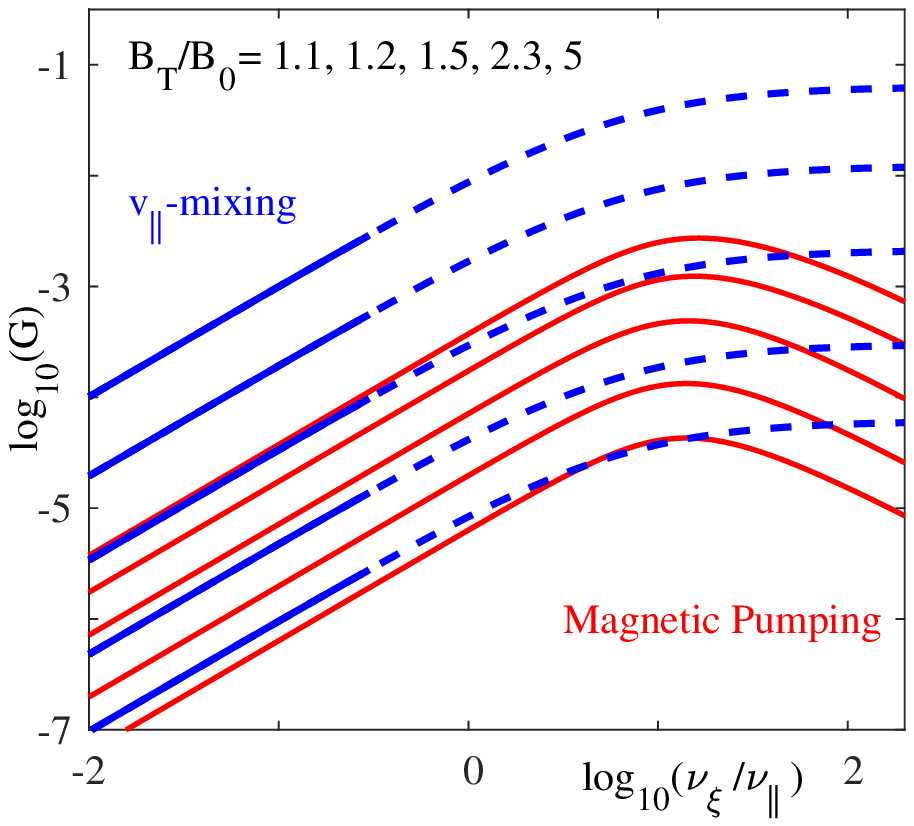}%
	\caption{Blue lines: The energization rate ${\cal G}$ by $\vz$-mixing as a function of $\nuC/\nuP$, calculated using Eq.~\eq{dg3} for $B_T/B_0 \in \{1.1, 1.2, 1.5, 2.3, 5 \}$. Indicated by full lines, the theory is expected to be valid for $\nuC/\nuP<1/3$.  
		 Red lines: For comparison the efficiency of magnetic pumping the red lines represent the similar ${\cal G}$  in Eq.~6 of Ref.~\cite{lichko:2020a}, evaluated with $\nu/f_{pump}=\nuC/\nuP$ and $C_K=1$, and considering  the  same magnetic perturbations as applied for the  $\vz$-mixing.
	}
	\label{fig:Pumpdouble}
\end{figure}
Above we introduced the 1D distribution $g=g(v,t)$ for characterizing the isotropic component of the background plasma. The main goal of the present section is to derive an evolution equation that describes the slow evolution of $g$. To accomplish this we need to consider the full 2D distribution, which we approximately describe as a  linear combinations of 
the fully $\vl$-mixed distribution $\aveP{g}$ and
the fully pitch angle scattered distribution $\aveC{g}$,  such that 
\begin{equation}
\hblabel{eq:FofG}
f = (1-\alpha)\aveC{g} + \alpha \aveP{g}\quad.
\end{equation}
The parameter $\alpha$ will be determined below and is dependent on the drive frequency, $\nuP$, of the parallel mixing compared to the  characteristic frequency, $\nuC$, of the pitch angle diffusion.

We further approximate the parallel mixing and pitch angle diffusion in terms of Krook-type operators, allowing us to 
 write the kinetic equation as 
\begin{equation}
\hblabel{eq:dfdtK}
\ddd{f}{t} = \nuP\left(\aveP{f} - f\right) + \nuC\left(\aveC{f} -f \right)\quad,
\end{equation}
which we through numerical analysis  (not included) find is a reasonable approximation for $\nuP \gtrsim  3\nuC$. Here $\nuP$ describes the characteristic frequency of the $\vz$-diffusion process, which will be on the order of the frequencies describing the magnetic perturbations. Similarly, $\nuC$ is the characteristic frequency of the pitch angle scattering process. 

Inserting Eq.~\eq{FofG} into Eq.~\eq{dfdtK}  yields
\begin{equation}
\hblabel{eq:dfdtK2}
\ddd{f}{t} = - K \left(\aveP{g} -  \aveC{g} \right)  +  \nuC \alpha\aveC{\delta g}\quad,
\end{equation}
with
\begin{equation}
\hblabel{eq:difA}
K= \nuC \alpha  -\nuP (1-\alpha)\,, 
\end{equation}
where we have used Eq.~\eq{pump} together with the rules 
\[
\aveC{\aveC{g}}=\aveC{g}
\,,\,\,
\aveP{\aveC{g}}= \aveP{g}
\,,\,\,
\aveP{\aveP{g}}=\aveP{g}\,.
\]
Note that the two first of these rules follow because $g(v)$ is isotropic such that $\aveC{g}=g$.

Next we use that direct differentiation of Eq.~\eq{FofG} with respect to time yields
\begin{equation}
\hblabel{eq:dfdtK3}
\ddd{f}{t} = \dot{\alpha}\left(\aveP{g} -\aveC{g} \right) + 
(1-\alpha)\aveC{\dot{g}} + \alpha \aveP{\dot{g}}\quad,
\end{equation}
where we used the notation $\dot{g}=\ddt{g}{t}$ and $\dot{\alpha}=\ddt{\alpha}{t}$. 
Matching the terms in Eqs.~\eq{dfdtK2} and \eq{dfdtK3} proportional to $\left(\aveP{g} -\aveC{g} \right)$ we find
\[
\dot{\alpha} = -K = \nuP (1-\alpha) - \nuC \alpha \quad. 
\]
This provides an evolution equation of the level of anisotropy parameterized by $\alpha$. At a timescale on the order of $1/\nuP$, the value of $\alpha$ will approach the steady state solution described by $K=0$ for which  
\begin{equation}
\hblabel{eq:alpha}
\alpha = \fd{\nuP}{\nuP+\nuC} \quad.
\end{equation}
Inserting Eq.~\eq{alpha} and $\dot{\alpha}=K=0$ into Eqs.~\eq{dfdtK2}  and \eq{dfdtK3} and equating the resulting  two expressions for $\ddt{f}{t}$ we find

\[
\fd{  \nuC  \nuP}{\nuP+\nuC} \aveC{\delta g} = \fd{\nuC}{\nuP+\nuC}\aveC{\dot{g}} +\fd{\nuP}{\nuP+\nuC} \aveP{\dot{g}}\quad.
\]
Then, by taking the $\aveC{...}$-average and using  the approximation that $\aveC{\aveP{\dot{g}}}
\simeq \aveC{\dot{g}}$ the right hand side simplifies and  we find
\begin{equation}
\hblabel{eq:dg1b}
\aveC{\dot{g}} = \fd{\nuC\nuP}{\nuP+\nuC}\aveC{\delta g}\quad.
\end{equation}
By inspection of Eq.~\eq{dfdtK} it becomes clear that the approximation $\aveC{\aveP{\dot{g}}}
\simeq \aveC{\dot{g}}$ above corresponds to the neglect of a second order time-derivative term of the approximate size $(1/\nuP)\partial \dot{g} /\partial t$. 

Given Eq.~\eq{dg1b}, we may now apply the form in Eq.~\eq{dg} to obtain the desired evolution equation for $g(v,t)$
\begin{equation}
\hblabel{eq:dg2}
\boxed{\ddd{g}{t} =
\frac{1}{v^2}\ddd{}{v} v^2 D \ddd{}{v} g\,,\quad
D=\nuP v^2 {\cal G} \,,}
\end{equation}
where 
\begin{equation}
\hblabel{eq:dg3}
{\cal G} = \fd{\nuC}{\nuP+\nuC  } \frac{h}{45 }
\left(\frac{h}{1+h} \right)^{3/2}\,,\quad
 h=\frac{B_T}{B_0} -1 \,.
\end{equation}

We have boxed Eq.~\eq{dg2} as it represents the main result of our analysis. 

\section{Discussion and Conclusion}

The expression for $\ddt{g}{t}$ in Eq.~\eq{dg2} has the form of   velocity diffusion, where the diffusion coefficient $D\propto \nuP v^2$ describes a process with a diffusive step-size proportional to velocity $\delta v\propto v$. 
Equivalently, the diffusive energy step-size is proportional to energy, as is characteristic of a Fermi-heating process. It is readily seen that a power-law distribution of the  form $g\propto v^{-\gamma}$ with $\gamma=3$, represents a steady state solution to Eq.~\eq{dg2}. For a more realistic representation of a physical system, particle sources and sinks can be added to Eq.~\eq{dg2}. In general, this will lead to power-law solutions with $\gamma > 3$ \cite{montag:2017}. The diffusion equation in Eq.~\eq{dg2} (as well as the similar form obtained in Ref.~\cite{lichko:2020a}) is therefore consistent with the power law distributions recorded {\sl in situ} by spacecraft through out the solar wind and the Earth's magnetosphere.

For wave dynamics with a typical magnitude $\tilde{dB} = dB/B_0$ we may approximate  $B_T\simeq B_0+dB$ and it follows that $h\simeq \tilde{dB}$. Given the dependency of  Eq.~\eq{dg2} on $h$, the efficiency of the energy diffusion for small $\tilde{dB}$ scales as $\tilde{dB}^{5/2}$, but falls off to a linear scaling for larger order unity wave amplitudes.  
The present model is obtained using the Krook-model in Eq.~\ref{eq:dfdtK}, which from  preliminary numerical results (not included here) is found to be a good approximation when the system is characterized by weak pitch angle scattering, $\nuC/\nuP \lesssim 1/3$. In Fig.~\ref{fig:Pumpdouble} the blue lines illustrate the predictions of Eq.~\eq{dg3} for ${\cal G}$, evaluated for the amplitudes of $B_T$ listed in the figure, and the full lines for  $\nuC/\nuP <  1/3$ correspond the range where the model is expected to be accurate.
For $\nuC/\nuP \gtrsim 1/3$ the Krook model in Eq.~\eq{dfdtK} becomes inaccurate because the pitch angle scattering will
cause the region-A and region-B electrons (see Fig.~\ref{fig:DoubleFig3}(b)) to mix without the separate $\mu \ddt{B}{t}$ heating/cooling of the two regions.  In fact, for $\nuC/\nuP \gtrsim 1$ we expect that heating by magnetic pumping will be  more efficient than heating by  $\vz$-mixing.

For comparison, the red lines in Fig.~\ref{fig:Pumpdouble} is obtained from the model of  magnetic pumping with trapped electrons developed in \cite{lichko:2020a}, where in Eq.~6 an expression is given for the form of  ${\cal G}_{pump}$ due to magnetic pumping, and we evaluate ${\cal G}_{pump}$ for the same magnetic perturbations as yielded the blue lines in Fig.~\ref{fig:Pumpdouble}. Furthermore, in this comparison  ${\cal G}_{pump}$ is obtained assuming $\nu/f_{pump} = \nuC/\nuP$. The model for ${\cal G}_{pump}$ also includes a factor,  $C_K$,  that calibrates the efficiency of a Krook scattering model to the efficiency of the Lorentz scattering operator. 
The Krook scattering model implemented here in Eq.~\eq{dfdtK} is equivalent to  $C_K=1$, and is thus the value used in calculating the red lines.

We observe that the predicted heating from $\vz$-mixing is up to two orders of magnitude larger than that expected from magnetic pumping. Physically, this result is reasonable because net energization in the magnetic pumping model requires pitch angle scattering during each pumping cycle. In contrast, the  $\vz$-mixing yields finite $\cEz$-energization   even if $\nuC=0$ (corresponding to the changes in the distributions between Figs.~\ref{fig:DoF2}(a) and \ref{fig:DoF2}(b)). As in Landau damping, for the limit of $\nuC=0$ the process is fully reversible for the hypothetical case where the mixing cycle is {\sl exactly} reversed. But given the fine scales structures that develops in velocity space after just a few mixing cycles, such are revesal is unlikely to occur in any physical system. As emphasized in \cite{lichko:2020a}, the electron energization is caused by mechanical work through the term $p_{\perp} \nabla \cdot \vv_{\perp}$ and is linked to the development of pressure anisotropy, which in the pumping model is  continuously being isotropized by pitch angle scattering. Meanwhile, for the $\vz$-mixing this anisotropy can build during multiple mixing cycles and  becomes more pronounced than the anisotropy that develop during a  single magnetic pump cycle.


The regime with $\nuC/\nuP \ll 1/3$ is likely to be relevant to the solar wind for which a recent analysis of the Strahl-electrons show that pitch angle scattering is mostly limited to the low level provided by Coulomb collisions between electrons and ions \cite{horaites:2019}. Meanwhile, for MMS bow-shock encounter analyzed in \cite{lichko:2020a} we estimate that $\nuC/\nuP \simeq  1/2$, whereas the analysis of a similar MMS bow-shock event \cite{amano:2020} infer much larger values of $\nuC/\nuP$. In future studies of {\sl in situ} spacecraft data, to help determine the relevant value of $\nuC/\nuP$ of a given dataset, we note that Eqs.~\eq{FofG} and \eq{alpha} can be fitted to electron data as provided by for example NASA’s Magnetospheric Multiscale (MMS) mission  \cite{Burch:2016a} and may prove useful for inferring $\nuC/\nuP$ directly from the observations.

The magnetic configurations considered here  are highly idealized. This adds to the need for developing new analytical and numerical techniques for simultaneously evaluating heating by both  $\vz$-mixing and magnetic pumping for more general magnetic perturbation geometries. Nevertheless, while the configurations considered are useful for providing physical insight to the heating mechanism, we expect that the energization theoretical rates obtained  will prove representative also for naturally occurring systems. This is emphasized by the result that the very  different scenarios considered in the main text and Appendix A, respectively, provide similar levels of $\vz$-mixing.

\section{Appendix A}
\begin{figure}
	\centering
	\includegraphics[width=13.8 cm]{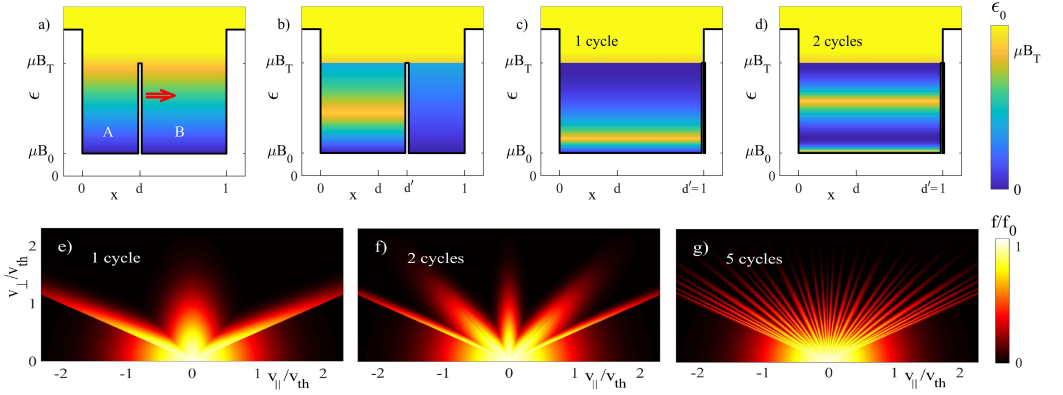}%
	\caption{Illustration of $\vz$-mixing by changing the location of a magnetic barrier initially located at $x=d$. In a-d) the color-contours represent $\cE_0$ as a function of $x$ and $\cE$, with the initial profile in a), while $b)$ and $c$ are computed with the barrier moved to $d'=0.6$ and $d'=1$, respectively. d) correspond to the result of two complete mixing cycles. e-g) Electron distributions computed using  Eq.~\eq{dE4} for 1, 2, and 5 complete mixing cycles, respectively. 
	}
	\label{fig:Compress}
\end{figure}

In this Appendix we consider a modified $\vz$-mixing scenario, which turns out to yield results very similar to those of Sections 2 and 3. As outlined in Fig.~\ref{fig:Compress}(a) we again consider a magnetic barrier at $x=d$ with height $\mu B_T$, separating the spatial dimension into region A and region B. We then examine the $\vz$-mixing that occurs as region A is expanded at the expense of region B, corresponding to the location of the barrier being moved from $d'=d$ towards $d'=1$. Requiring again that the parallel action integrals be conserved $(J_A=J_{A0})$, it follows that the region A electrons are being cooled with
\begin{equation}
\label{eq:rA}
\cE_0=\left(\fd{d'}{d}\right)^2\left(\cE-\mu B_0\right) + \mu B_0\quad.
\end{equation}
Meanwhile, the region B electrons are being heated and the initial energy $\cE_0$ and present energy $\cE$ are similarly described by \begin{equation}
\label{eq:rB}
\cE_0=\left(\fd{1-d'}{1-d}\right)^2\left(\cE-\mu B_0\right) + \mu B_0\quad.
\end{equation}

As region B is contracting the electrons confined to this region are all subject to $\vz$ heating and will eventually reach the energy $\cE=\mu B_T$ where they can overcome the barrier. After clearing the barrier they will immediately experience the cooling of region A and will therefore become trapped in region A.  For an initial value of $\cE_0$ we obtain from Eq.~\eq{rB} the value of $d'=d_T$ when this transition occurs 
\begin{equation}
\label{eq:dT}
\left(1-d_T\right)^2= (1-d)^2 \left(\fd{\cE_0-\mu B_0}{\mu(B_T-B_0)}\right)\quad.
\end{equation}

To further established the relationship between $\cE_0$ and $\cE$ after the transition, we characterise the subsequent cooling in region A  for $d_T\leq d'\leq1$. Considering Eq.~\eq{rA}, it follows that this cooling must be governed by  
\begin{equation}
\label{eq:rBA}
\mu B_T=\left(\fd{d'}{d_T}\right)^2\left(\cE-\mu B_0\right) + \mu B_0\quad.
\end{equation}
Combining Eqs.~\eq{dT} and \eq{rBA} by eliminating $d_T$ while solving for $\cE_0$ we obtain 
\begin{equation}
\label{eq:rBA2}
\cE_0=\fd{\mu(B_T-B_0)}{(1-d)^2}\left(1-d'\,\sqrt{\fd{\cE-\mu B_0}{\mu(B_T-B_0)}}\,\right)^2+\mu B_0\quad.
\end{equation}
We further introduce the transition energy 
\[
\cE_T= \fd{d^2}{{d'}^2} \,\mu(B_T-B_0) + \mu B_0\quad,
\]
obtained by solving Eq.~\eq{rA} for $\cE$ with $\cE_0=\mu B_T$. The electrons in region A are then characterized by Eq.~\eq{rA} for $\mu B_0 \leq \cE \leq \cE_T$, and by Eq.~\eq{rBA2} for
 $\cE_T \leq \cE \leq \mu B_T$. The electrons in region B are characterized by 
 Eq.~\eq{rB} for the full interval $\mu B_0 \leq \cE \leq \mu B_T$. With the initial barrier at $d=0.4$, the derived relationship between $\cE$ and $\cE_0$ is illustrated in Figs.~\ref{fig:Compress}(b,c) evaluated for $d'=0.6$ and $d'=1$, respectively. Note that electrons with $\cE > \mu B_T$ are not affected by the changes in the location of the magnetic barrier. 

In the present scenario, a mixing cycle is complete when $d'=1$ and all electrons are then characterized by the region-A expressions. Similar to the derivation in Section 2, we readily obtain recurrence relations for the impact of $N$ complete mixing cycles:
\begin{equation}
\label{eq:dE4}
\cE_{N-1} =  \begin{cases} \,\, \fd{\cE_N-\mu B_0}{d^2} +\mu B_0  & \mbox{for}\quad \mu B_0< \cE_N <\cE_T
\\[3ex]
\,\,  \fd{\mu(B_T- B_0)}{(1-d)^2}\left(1-\sqrt{\fd{\cE_N-\mu B_0}{\mu(B_T-B_0)}}\,\right)^2
+ \mu B_0
 & \mbox{for}\quad  \cE_T<\cE< \mu B_T 
\\[3ex]
\,\,\cE_N & \mbox{for}\quad  \mu B_T< \cE_N
\end{cases} \,\,,
\end{equation}
and because $d'=1$ the transition energy is here characterized by 
\[
\cE_T= d^2\mu(B_T-B_0) + \mu B_0\quad. 
\]

In  Fig.~\ref{fig:Compress}(d) we display the predictions of Eq.~\eq{dE4} computed for 2 mixing cycles, and similar to the distributions in Section 3,  Fig.~\ref{fig:Compress}(e-g) display the distributions that result after 1, 2, and 5 cycles. Although the mixing process here is different from that of Sections 2 and 3, the final result is again a rapid $\vz$ mixing and diffusion for the magnetically trapped electrons.

\section{Appendix B}

We will here derive the expression for $\delta g$ given in Eq.~\eq{dg}. For this we apply the procedure outlined in Fig.~\ref{fig:DoF2} imposing particle conservation between $\aveC{g}$, $\aveP{g}$ and $\aveC{\aveP{g}}$ for the differential velocity region encircled by the cyan and magenta lines, respectively. 
In our analysis we will consider the distributions on the form  $f(\cEz,\cEp)$ normalized such that $n=\int f(\cEz,\cEp) d^3v$.
Because
\[
d\vz = \fd{d\cEz}{m\vz}=\fd{d\cEz}{\sqrt{2m\cEz}}\quad,\qquad
2\pi\vp d\vp = \fd{2\pi}{m} d\cEp
\]
we have
\begin{equation}
\hblabel{eq:n}
n= \fd{\pi\sqrt{2}}{m^{3/2}}\int\int f(\cEz,\cEp) \fd{1}{\cEz^{1/2}}\, d\cEz d\cEp\quad.
\end{equation}

\newcommand{\fp}{{f_\perp}}

First, the background distribution  $\aveC{g}$ is isotropic, but during the mixing process  rapid diffusion occur in $\cEz$ for all the electrons trapped by $B_T$. Again, the trapped electrons are those with $\cEz< h \cEp$, where $h = B_T/B_0-1$, and within this fully diffused velocity region of  $\aveP{g}$ is independent of $\cEz$; we will characterize this part of the distribution as $\fp(\cEp)$, i.e. $\fp(\cEp)= \aveP{g}$ for $\cEz< h \cEp$.  From Eq.~\ref{eq:n}, particle conservation for the differential velocity regions outlined in cyan in Figs.~\ref{fig:DoF2}(a,b) then imposes that 
\[
\Delta\cEp \int_0^{h\cEp}\fp(\cEp) \fd{1}{\cEz^{1/2}}\, d\cEz
=\Delta\cEp \int_0^{h\cEp} g(\cE)  \fd{1}{\cEz^{1/2}}\, d\cEz
\]
or
\[
\fp(\cEp) = \fd{1}{2(h\cEp)^{1/2}}
\int_0^{h\cEp} g(\cE)  \fd{1}{\cEz^{1/2}}\, d\cEz\quad.
\]
For approximate evaluation of this integral we use that $\cE=\cEz+\cEp$ and Taylor expand $g$ about $\cEp$, such that 
\[
g(\cE) \simeq  g(\cEp) + g^\prime(\cEp)\cEz + \frac{1}{2} g^{\prime\prime}(\cEp)\cEz^2\quad,
\]
and it then follows that 
\begin{eqnarray}
\hblabel{eq:fp}
\fp(\cEp) &\simeq& \fd{1}{2(h\cEp)^{1/2}}\,\left[ 2g \cEz^{1/2} +\frac{2}{3} g^{\prime}  \cEz^{3/2}+  \frac{1}{5} g^{\prime\prime}  \cEz^{5/2}
\right]_0^{h\cEp}\,,
\nonumber\\[2ex]
&\simeq&g(\cEp) +\frac{1}{3}h\cEp g^{\prime} (\cEp)
+\frac{1}{10}(h\cEp)^2 g^{\prime\prime} (\cEp)\,\,.
\end{eqnarray}

For what comes next, Taylor expansion of $\fp(\cEp)$ to second order becomes useful
\begin{equation}
\fp(\cEp)
 \simeq \fp(\cE) -  \,\fp^\prime(\cE)  \cEz + \frac{1}{2} \fp^{\prime\prime}(\cE)\cEz^2
 \,.
 \nonumber
\end{equation}
We then use Eq.~\ref{eq:fp} to obtain expressions for the derivatives of $\fp$ such that
\[
\fp^\prime(\cE) \simeq
\left(1+\frac{h}{3} \right) g^{\prime}(\cE) + \frac{h}{3}\cE g^{\prime\prime}(\cE))
+\frac{1}{5}h^2\cE g^{\prime\prime} (\cE)\quad,
\]
and
\[
\fp^{\prime\prime}(\cE) \simeq
\left(1+\frac{2h}{3} +\frac{h^2}{5} \right) g^{\prime\prime}(\cE)\quad.
\]
Using the cosine-pitch-angle variable introduced above we have $\cEz=\xi^2\cE$, and it follows that 
\begin{eqnarray}
\hblabel{eq:fp2}
\fp(\cEp)&\simeq\,\,&
g(\cE) +\frac{1}{3}h\cE g^{\prime} (\cE)
+\frac{1}{10}(h\cE)^2 g^{\prime\prime} (\cE)
\nonumber \\[1ex]
&^-&\xi^2\cE \left(
\left(1+\frac{h}{3} \right) g^{\prime}(\cE) + \left( \frac{h}{3}
+\frac{h^2}{5}\right)\cE g^{\prime\prime} (\cE)\right)
\nonumber \\[1ex]
&+&
\frac{\xi^4}{2}\cE^2 \left(1+\frac{2h}{3} +\frac{h^2}{5} \right) g^{\prime\prime}(\cE)\quad.
\end{eqnarray}



With Eq.~\ref{eq:fp2} we now have an expression for the diffused region of $\aveP{g}$ in terms of $g$, readily evaluated as a function of $\cE$ and $\xi$. To continue and obtain an expression for $ \delta g= \aveC{\aveP{g}-\aveC{g}}  $ we apply that the number of particles in the differential speed intervals encircled in magenta in Figs.~\ref{fig:DoF2}(b,c) must be identical. In general, the number of particles in an interval $dv$ can be computed as $4\pi v^2dv\int_0^1 f(v,\xi) d\xi$. Meanwhile, $\aveP{g}$ and $\aveC{g}$ only differ in the trapped region characterized by $\xi\leq k$, where  $ k^2=h/(1+h)$.
It then follows that  
\[\delta g \int_0^{1} d\xi = \int_0^{k} (\fp(\cEp) -g(\cE) ) d\xi\quad.
\]
Here, of course, $\int_0^1 d\xi=1$, and  we proceed to evaluate directly the right-hand-side  using Eq.~\ref{eq:fp2} well suited for the required integration over $\xi$ at constant $\cE$:
\begin{eqnarray}
\delta g&\simeq\,\,&
\frac{1}{3}kh\cE g^{\prime} (\cE)
+\frac{1}{10}k(h\cE)^2 g^{\prime\prime} (\cE)
\nonumber \\[1ex]
&\,\,-& \frac{1}{3} k^3\cE \left(
\left(1+\frac{h}{3} \right) g^{\prime}(\cE) + \left(\frac{h}{3}
+\frac{h^2}{5}\right)\cE g^{\prime\prime} (\cE)\right)
\nonumber \\[1ex]
&\,\,+&
\frac{1}{10}k^5\cE^2\left( \left(1+\frac{2h}{3} +\frac{h^2}{5} \right) g^{\prime\prime}(\cE)\right)\quad,
\nonumber
\end{eqnarray}
This expression has the form
\begin{equation}
\hblabel{eq:delf}
\delta g\simeq A \cE g^{\prime} (\cE) + B\cE^2 g^{\prime\prime} (\cE)\quad,
\end{equation}
with
\[
\frac{A}{k}=\frac{1}{3}h
-\frac{1}{3} k^2
\left(1+\frac{h}{3} \right)\quad,
\]
and
\[
\frac{B}{k}=
\frac{1}{10}h^2 -\frac{1}{9} k^2h
-\frac{1}{15} k^2h^2+\frac{1}{10}k^4 \left(1+\frac{2h}{3} +\frac{h^2}{5} \right)\,.
\]
Using $k^2=h/(1+h)$, the expressions for $A$ and $B$ reduce to 
\[A=\frac{2}{9} h\left(\frac{h}{1+h}\right)^{3/2}\,\,,
\quad
B=\frac{2}{5} A  + \frac{4}{75} h^2 \left(\frac{h}{1+h}\right)^{5/2} \quad.
\]
We further notice that Eq.~\eq{delf} can also be written as
\[
\delta g\simeq A \frac{v}{2}  \ddd{g}{v} + B\left(\frac{-v}{4}
\ddd{g}{v} + \frac{v^2}{4} \ddd{^2g}{v^2}\right)\quad.
\]
To within the order and accuracy of the applied Taylor expansions we  have $B=2A/5$ and the result stated in  Eq.~\eq{dg} then follows from simple manipulations:
\[
\delta g \simeq
\frac{A}{10} \frac{1}{v^2}\ddd{}{v} v^4 \ddd{}{v} g \quad.
\]


\end{document}